\documentclass[12pt]{article}
\begin{document}
\begin {titlepage}
\begin{flushright} ULB-TH/02-10\\ DCPT-02/17\\
hep-th/0203098\\
\end{flushright}
\vskip 2cm

\begin{center} {\large \bf  The Bosonic Ancestor of Closed and Open
Fermionic Strings}
\footnote{Contribution to the Francqui meeting  ``Strings and
Gravity: Tying the Forces Together'', Brussels, October 21-23,
2001. Presented by F. Englert at the  Corfu Summer Institute on
Elementary Particle Physics, August 31 - September 20, 2001. }\\
\vspace{.5cm}  Fran\c cois Englert,${}^{a,}$\footnote{ E-mail :
fenglert@ulb.ac.be} Laurent Houart${}^{b,}$\footnote{ Research Associate of 
the FNRS, E-mail : lhouart@ulb.ac.be} and Anne Taormina${}^{c,}$\footnote{
E-mail : anne.taormina@durham.ac.uk}\\
\vspace{1cm} ${}^a${\it Service de Physique Th\'eorique}\\ {\it
Universit\'e Libre de Bruxelles, Campus Plaine, C.P.225}\\ {\it
Boulevard du Triomphe, B-1050 Bruxelles, Belgium}\\
\vspace{.2cm}
 ${}^b${\it Physique Th\'eorique et Math\'ematique }\\  
{\it Universit\'e Libre de Bruxelles, Campus Plaine C.P. 231}\\ 
{\it Boulevard du Triomphe, B-1050 Bruxelles, Belgium.}\\
\vspace{.2cm}
 ${}^c${\it Department of Mathematical Sciences }\\  {\it University
of Durham}\\ {\it South Road, DH1 3LE Durham, England}\\
\end{center}
\vskip 1cm
\begin{abstract}

 We review the emergence of the ten-dimensional fermionic closed
string theories from subspaces of the Hilbert space  of the
26-dimensional  bosonic closed string theory compactified on an
$E_8\times SO(16)$ lattice. They arise from a consistent truncation
procedure which generates    space-time fermions out of bosons. This
procedure  is   extended to  open string sectors. We prove, from  bosonic 
considerations alone,  that
truncation  of the unique tadpole-free $SO(2^{13})$   bosonic string
theory compactified on the above lattice determines  the anomaly free
Chan-Paton group of the  Type I theory.  It also yields  the 
Chan-Paton groups making Type~O  theories tadpole-free. These results
establish a link between all M-theory strings and the bosonic string within
the framework of conformal field theory. Its
  significance is discussed.
\end{abstract}
\end {titlepage}
\addtocounter{footnote}{-4}
\section{Introduction}

\parskip 11pt plus 1pt
\setlength{\parindent}{0cm}

It has been demonstrated previously that all the ten-dimensional
closed superstring theories (Type IIA, Type IIB and the two distinct
heterotic superstrings) are hidden in the Hilbert space of the
26-dimensional closed bosonic string theory
\cite{cent}.  The emergence of space-time fermions and of
supersymmetry, anticipated by Freund
\cite{freund}, is an impressive property of the bosonic string. The
generation of space-time fermions out of bosons appears in reference
\cite{cent} as a stringy generalisation of the field theoretical
mechanism by which non abelian monopoles become fermions in an
appropriate environment~\cite{thooft}.

The superstrings arise from a  toroidal
compactification of the bosonic strings on an $E_8\times E_8$ lattice
where only the second
$E_8$ plays an active r\^ole. The superstring content of the bosonic
string  appears when all states pertaining to the first $E_8$ are 
removed from the spectrum. We call this
 a ``truncation''. The truncation required to reveal  superstrings
has in fact to be extended to all oscillator states in four of the
eight compact dimensions spanned by the second $E_8$ lattice of zero
modes. However, some particular zero modes  have to be retained in
these dimensions. They  play an essential r\^ole in the construction.

The  $E_8$ lattice is a sublattice of the
$SO(16)$ lattice and the theory is more elegantly formulated in terms
of
$E_8\times SO(16)$ \cite{ens}. This formulation was in fact a crucial
step, because it led to uncover not only  the  superstrings,  but
also the non-supersymmetric fermionic strings\footnote{We shall call
fermionic strings all strings containing space-time fermions in
separate left or right sectors whether or not they are projected out
in the closed string spectrum.}. Type OA and Type OB \cite{sw} and all
the consistent non-supersymmetric ten-dimensional heterotic strings
discovered in reference
\cite{harveydix} emerged indeed from the  bosonic
string compactified on an $E_8\times SO(16)$ lattice, using the same
truncation  that generated the superstrings out of its
$E_8\times E_8$ sublattice
\cite{lls,y,ns}.

We  extend the truncation process to open
string sectors.  We   find that the
Chan-Paton group  $SO(32)$ which is required to make the
supersymmetric Type I superstring anomaly free, as well as the
Chan-Paton groups eliminating tadpoles in Type O fermionic strings, are
determined by  the bosonic parent. The  significance of these
results will be
discussed.

This report summarises the main results obtained in reference \cite{eht}.

\newpage

\section{Truncation of the closed bosonic  string}

\subsection{Space-time fermions from bosons}

To accommodate space-time fermions   in
 the  26-dimensional bosonic string one must meet three requirements:

\hskip 1cm a) A continuum of bosonic zero modes must be removed. This
can be achieved by compactifying  $d=24-s$ transverse dimensions  on a
$d$-dimensional torus. This leaves $s+2$ non-compact dimensions with
transverse group
$SO_{trans}(s)$.

\hskip 1cm b) Compactification must generate an internal group
$SO_{int} (s)$ admitting spinor representations\footnote{We shall 
designate all locally isomorphic groups by the same notation.}. 
This can be achieved by toroidal compactification on the Lie lattice of a
simply laced Lie group
${\cal G}$ of rank $d$ containing a subgroup $SO_{int} (s)$. The
latter is then mapped onto  $SO_{trans} (s)$  in such a way that the
diagonal algebra
$ so_{diag}(s) ={\rm diag} [so_{trans}(s) \times so_{int}(s)]$ becomes
identified with a new transverse algebra. In this way, the spinor
representations of  $SO_{int} (s)$ describe  fermionic states because
a rotation in space induces a half-angle rotation on these states. 

\hskip 1cm c)  The consistency of the above procedure relies on the
possibility of extending the diagonal algebra $so_{diag}(s)$ to the
new full Lorentz algebra
$so_{diag}(s+1,1)$,  a highly non trivial constraint. To break the
original Lorentz group $SO(25,1)$ in favour of the new one,   a
truncation consistent with conformal invariance must be performed on
the physical spectrum of the bosonic string.  Actually, states
described by 12 compactified bosonic fields must be truncated, except
for zero modes \cite{cent,ens}. 

This  can be understood heuristically by
counting the missing central charge, hidden in the light cone gauge,
needed for the  superghost and the longitudinal and time-like Majorana
fermions.  In units where each
two-dimensional boson contributes $1$ to the Virasoro central  charge,
the  superghost contributes $11$, while the longitudinal and
time-like Majorana fermions contribute $2 \times 1/2$.  Therefore, one
must truncate 12 compactified bosonic fields. The larger value of the
transverse dimension is then $s=8$, as the four remaining
dimensions in the light cone gauge can then accomodate the
internal group $SO_{int}(8)$.  The     highest available space-time
dimension accommodating fermions is therefore
$s+2=10$.  The truncation should also remove the zero-point energies of
the  truncated oscillators. We shall show below that this amounts to
retain some zero-modes of the  12 truncated bosonic fields. 

\subsection{Compactification and truncation}
 Consider the bosonic closed string compactified on a
$d$-dimensional torus. In terms of the left and right compactified
momenta, the mass spectrum is
\begin{eqnarray}
\label{spectrum} {\alpha^\prime m_L^2\over 4}&=&  \alpha^\prime {\bf
p}_L^2 + N_L -1\ , \nonumber \\ {\alpha^\prime m_R^2\over 4}&=&
\alpha^\prime {\bf p}_R^2 + N_R -1 \ , \\ &\hbox{and}&\nonumber\\ m^2
= { m_L^2\over 2}+{ m_R^2\over 2}\quad &;&\quad m_L^2=m_R^2\ .
\end{eqnarray}
 In Eq.(\ref{spectrum})
$N_L$ and
$N_R$ are the oscillator numbers in 26-dimensions and the zero modes
$\sqrt{2\alpha^\prime}{\bf p}_L$,
$\sqrt{2\alpha^\prime}{\bf p}_R$ span a
$2d$-dimensional even self-dual Lorentzian lattice with negative
(resp. positive) signature for left (resp. right) momenta. This
ensures modular invariance of the closed string spectrum 
\cite{narain}. For generic toroidal compactifications, the massless
vectors
$\alpha_{-1,R}^\mu ~\alpha_{-1,L}^i\vert 0_L,0_R\rangle$ and
$\alpha_{-1,L}^\mu ~\alpha_{-1,R}^i \vert 0_L,0_R\rangle$, where the
indices
$\mu$ and
$i$ respectively refer to non-compact and  compact dimensions,
generate a local symmetry  $[U_L(1)]^d\times [U_R(1)]^d$. But more
massless vectors arise when $ \sqrt{2\alpha^\prime}{\bf p}_L$ and
$\sqrt{2\alpha^\prime}{\bf p}_R$ are  roots of simply laced groups
${\cal G}_L$ and ${\cal G}_R$ (with roots length $\sqrt{2}$). The
gauge symmetry is  enlarged to  ${\cal G}_L\times {\cal G}_R$  and the
theory is modular invariant provided the lattice of zero modes  is 
self-dual Lorentzian even. In particular, this is the case for
compactification on  a
${\cal G}\times {\cal G}$ lattice where
${\cal G}$ is any semi-simple simply laced group  of rank $d$, if both
$\sqrt{2\alpha^\prime}{\bf p}_L$ and $\sqrt{2\alpha^\prime}{\bf p}_R$
span the full weight lattice
$\Lambda_{weight}$ of
${\cal G}$,  but are constrained to be in the same conjugacy class.
Namely
 $\sqrt{2\alpha^\prime}({\bf p}_L-{\bf p}_R)$ must be on the root
lattice
$\Lambda_{root}$ of ${\cal G}$ \cite{en}.  Hereafter, such lattice
will be referred  to as the EN lattice of ${\cal G}$ .

Now perform  a toroidal compactification on  a 16-dimensional 
${\cal G}_L\times {\cal G}_R$ lattice and take 
${\cal G}_R \supset SO_{int}(8)$. Then truncate the spectrum of the right
sector of the theory, by tentatively removing {\em all} states (i.e.
oscillators and momenta) in $12$  compactified dimensions keeping only the
$SO_{int}(8)$.

The centre of the covering group of $SO(8)$ is  $Z_2\times Z_2$.
Its   four elements partition the weight lattice in  four conjugacy
classes
$(o)_8,(v)_8, (s)_8, (c)_8$  isomorphic to the root lattice. The
$(o)_8$ lattice is the root lattice itself and contains the element  $
\sqrt{2\alpha^\prime}{\bf p_o}=(0,0,0,0)$. The
$(v)_8$ lattice is the vector lattice whose smallest weights are eight
vectors of norm one; in an orthonormal basis, these are
$\sqrt{2\alpha^\prime}{\bf p_v}=(\pm1, 0,0,0)$ + permutations. The
$(s)_8$ and $(c)_8$ lattices are spinor lattices whose smallest
weights also have norm one and are the eightfold degenerate vectors
$\sqrt{2\alpha^\prime}{\bf p_{s,c}}=(\pm1/2,\pm1/2,\pm1/2,\pm1/2)$
with even (for class $(s)_8$) or odd (for class $ (c)_8$) number of
minus signs. The structure of the weight lattice of all $SO(4m)$
groups in a
$2m$-dimensional Cartesian basis is the same.

One must however keep zero modes in the 16 compact
dimensions  in such a way that
\begin{equation}
\label{ghost} \alpha^\prime{\bf p}_R^2[{\cal G}_R] = \alpha^\prime{\bf
p}_R^2[SO(8)] +{1\over 2}\ .
\end{equation}
These zero modes are required to remove the
zero-point energy (-12/24) contribution to the energy of the states
taken out by the truncation.

As shown below, Eq.(\ref{ghost}) can be satisfied by choosing
\begin{equation}
\label{group} {\cal G}_R= E_8 \times SO(16)\ ,
\end{equation} and we shall prove at the end of section {\bf 2.3}
that from this choice  all  known ten-dimensional closed fermionic
strings emerge from the same truncation scheme.  We decompose
$SO(16)$  in $SO^{\,
\prime}(8) \times SO(8)$ and truncate all  states created  by
oscillators in the 12 dimensions defined by the $E_8\times
SO^{\,\prime}(8)$ root lattice. To satisfy  Eq.(\ref{ghost}), we keep
only  $SO(16)$ zero modes which are
$SO^{\,\prime}(8)$ vectors of norm one. The latter are  chosen as
follows.

The decomposition of an $SO(16)$ lattice in terms of
$SO^{\,
\prime}(8)
\times SO(8)$ lattices yields
\begin{eqnarray} (o)_{16} = [(o)_{8^\prime}\oplus (o)_{8}] &+&
[(v)_{8^\prime}\oplus (v)_{8}] \ ,\nonumber\\ (v)_{16} =
[(v)_{8^\prime}\oplus (o)_{8}] &+& [(o)_{8^\prime}\oplus
(v)_{8}]\nonumber\ ,\\ (s)_{16} = [(s)_{8^\prime}\oplus (s)_{8}]& +&[
(c)_{8^\prime}\oplus (c)_{8}]\nonumber\ ,\\ (c)_{16} =
[(s)_{8^\prime}\oplus (c)_{8}] &+& [(c)_{8^\prime}\oplus (s)_{8}]\ .
\end{eqnarray} The vectors of norm one in $SO^{\, \prime}(8)$ are the
4-vectors $\sqrt{2\alpha^\prime}{\bf p^\prime_v},
\sqrt{2\alpha^\prime}{\bf p^\prime_s}$ and
$\sqrt{2\alpha^\prime}{\bf p^\prime_c}$ described above. We choose one
vector
$\sqrt{2\alpha^\prime}{\bf p^\prime_v}$ and one vector
$\sqrt{2\alpha^\prime}{\bf p^\prime_s}$ (or equivalently
$\sqrt{2\alpha^\prime}{\bf p^\prime_c}$).

This gives the truncations
\begin{eqnarray}
\label{truncations} &&(o)_{16} \rightarrow   (v)_{8} \qquad (v)_{16}
\rightarrow   (o)_{8}\ ,\nonumber\\ &&(s)_{16} \rightarrow   (s)_{8}
\qquad (c)_{16} \rightarrow   (c)_{8} \ ,\\
\hbox{or}\nonumber \\ \label{truncationc} &&(o)_{16} \rightarrow  
(v)_{8}
\qquad (v)_{16}
\rightarrow   (o)_{8}\ ,\nonumber\\ &&(s)_{16} \rightarrow   (c)_{8}
\qquad (c)_{16} \rightarrow   (s)_{8} \ .
\end{eqnarray} It follows from the closure of the Lorentz algebra that
states belonging to the lattices $(v)_{8}$ or $(o)_{8}$   are bosons
while those belonging to the spinor lattices $(s)_{8}$ and $(c)_{8}$
are space-time fermions. 

It is easily verified that the choice of
zero modes
$\sqrt{2\alpha^\prime}{\bf p^\prime_v}$ and
$\sqrt{2\alpha^\prime}{\bf p^\prime_s}$ (or $\sqrt{2\alpha^\prime}{\bf
p^\prime_c}$)  preserve modular invariance in the truncation
provided  one flips the sign of the lattice partition functions of the lattices
$(s)$ and
$(c)$, in accordance with the spin-statistic theorem~\cite{eht}\footnote{For
an alternate proof of the modular invariance of the trucated theory, see
ref.\cite{ns}.  Generalisation to multiloops is given in ref. \cite{s}.}.

\subsection{Closed fermionic strings}

We now explain how the  consistent closed fermionic ten-dimensional strings
\cite{sw, harveydix} are   obtained by truncation from the
26-dimensional bosonic string.

First we consider the fermionic theory emerging from truncation in
both left and right sectors. We thus examine a compactification on
both sectors with
${\cal G}_L={\cal G}_R=E_8\times SO(16)$. In the truncated theory,
$(E_8)_{L,R}$ merely disappear, and we therefore only discuss in
detail the fate of the $SO(16)_{L,R}$ representations under
truncation. In the full bosonic string,  the representations of
$SO(16)$ entering the lattice partition function are restricted by modular
invariance\footnote{Strictly speaking the modular invariance does not
fully determine the representations of $SO(16)$ at this level (because
$\gamma(s)=\gamma(c)$). However at the level of the amplitudes, this
ambiguity is lifted.}.  Typically, such partition function is a sum of
products of left and right partition functions
$\bar\gamma_{\alpha L},\gamma_{\alpha R}~,~[\, \alpha = (o),(v),(s),(c)\, ]$.

Using the modular transformation properties, one finds
that there are two distinct modular invariant partition functions.
The first one is,
\begin{eqnarray}
\label{first} (\bar\gamma_{(o)_{16},L}+\bar\gamma_{(s)_{16},L})
(\gamma_{(o)_{16},R}+\gamma_{(s)_{16},R})\nonumber\\ &&\hskip
-8cm=\bar\gamma_{(o)_{16},L}\gamma_{(o)_{16},R}+
\bar\gamma_{(s)_{16},L}\gamma_{(o)_{16},R} +\bar\gamma_{(o)_{16},L}
\gamma_{(s)_{16},R}+\bar\gamma_{(s)_{16},L}\gamma_{(s)_{16},R}\ ,
\end{eqnarray} and the second is,
\begin{equation}
\label{second}
\bar\gamma_{(o)_{16},L}\gamma_{(o)_{16},R}+\bar\gamma_{(v)_{16},L}
\gamma_{(v)_{16},R}+
\bar\gamma_{(s)_{16},L}\gamma_{(s)_{16},R}+\bar\gamma_{(c)_{16},L}\gamma_{(c)_{1
6},R}\ .
\end{equation}
 Eq.(\ref{first}) can be rewritten as
$\bar\gamma_{(o)_{E8},L}\gamma_{(o)_{E8},R} $ where the subscript
$(o)$ refers to the  root lattice of  $E_8$ which is a sublattice of
the
$SO(16)$ weight lattice.  Eqs.(\ref{first}) and (\ref{second})
describe compactifications on the EN lattices  of   $ E_8\times E_8$ 
and of  $ E_8\times SO(16)$.

 To interpret the result of the truncation in conventional terms, we
note that the $SO(8)$ partition  functions
$\gamma_{(o)_{8}}$ and $\gamma_{(v)_{8}}$, divided by the Dedekind
functions arising from the bosonic states in the eight  non compact
transverse dimensions are  the Neveu-Schwarz partition functions with
the `wrong' and `right' GSO projection $(NS)_-$ and
$(NS)_+$.  The partition  functions
$\gamma_{(s)_8}$ and $\gamma_{(c)_8}$, divided  by the same Dedekind
functions,  form the two Ramond partition functions of opposite
chirality
$R_+$ and
$R_-$.

Choosing the ghosts
$\sqrt{2\alpha^\prime}{\bf p^\prime_v}$  and
$\sqrt{2\alpha^\prime}{\bf p^\prime_s}$ in both sector, and
truncating in accordance with Eq.(\ref{truncations}), the first
partition function Eq.(\ref{first}) yields the supersymmetric chiral
closed string,
\begin{equation}
\label{twob}
 {\bf IIB}: \quad (NS)_+\ (NS)_+ + R_+ \ (NS)_+ + (NS)_+ \ R_+ + R_+\
R_+ \ .
\end{equation} Replacing  ${\bf p^\prime_s}$ by  ${\bf p^\prime_c}$
in, say, the right sector we get, using Eq.(\ref{truncationc}), the
non chiral supersymmetric closed string
\begin{equation}
\label{twoa}
 {\bf IIA}: \quad (NS)_+\ (NS)_+ + R_+ \ (NS)_+ + (NS)_+ \ R_- + R_+\
R_-\ .
\end{equation} The same choices of the ghosts in the second partition
function yield the following non-supersymmetric strings
\begin{equation}
\label{ob}
 {\bf 0B}: \quad (NS)_+\ (NS)_+ +(NS)_-\ (NS)_- + R_+\ R_++ R_- \ R_-
\ ,
\end{equation}
\begin{equation}
\label{oa}
 {\bf 0A}: \quad (NS)_+\ (NS)_+ +(NS)_-\ (NS)_- + R_+\ R_-+ R_- \ R_+
\ .
\end{equation}

 Heterotic strings are generically obtained from compactification on
${\cal G}_L\times {\cal G}_R$ by only truncating in the right sector
with
${\cal G}_R= E_8 \times SO(16)$. The partition
function constructed on  ${\cal G}_L\times [E_8 \times SO(16)]_R$
must  be modular invariant. We may replace the Lorentzian metric by a
Euclidean one and drop the $E_8$ to preserve invariance under
translation
$(\tau \to \tau +1)$ in the Euclidean metric. This reduces the
problem of finding all heterotic strings obtainable in this way  to
that of finding all 24-dimensional Euclidean even self-dual lattices
containing a sublattice
$\Lambda(SO(16))$. All 24-dimensional even self-dual  Euclidean 
lattices have been classified: they are known as the  Niemeier
lattices. Heterotic strings are obtained from the relevant Niemeier
lattices by the truncation Eq.(\ref{truncations}) (or equivalently,
by Eq.(\ref{truncationc})) \cite{lls}. 

\section{ Brane fusion and the open string sectors}
\subsection{The bosonic open string ancestor}

We first review the derivation of the existence of  a 26-dimensional  
open bosonic string  free of massless tadpole divergences \cite{sob}.

The Chan-Paton group of the 26-dimensional unoriented,
uncompactified, open string theory may be fully determined by the 
tadpole condition \cite{sagn1,cal,cai}. The full one-loop vacuum
amplitude  of a theory with open and closed unoriented strings
comprises the four loop amplitudes with vanishing Euler
characteristic: the torus ${\cal T }$, the Klein bottle
${\cal K }$, the annulus ${\cal A}$ and the M{\"o}bius strip ${\cal
M}$. The last three amplitudes contain ultraviolet divergences which
are conveniently analysed in  the transverse channel.  This  channel
describes the tree level exchange  of zero momentum closed string
modes between holes and/or crosscaps.   The divergences appear there
in the infrared limit and  are associated with the exchange of
tachyonic and massless modes.  One ensures the  tadpole condition,
that is the cancellation of the  divergences due to the massless
modes in the total amplitude, by fixing the Chan-Paton group. Here,
the divergence is related  to  the  dilaton tadpole, a non-zero one
point function of a closed vertex operator on the disk or on the
projective plane.  If the tadpole condition is not imposed,  the
low-energy  effective action acquires a dilaton potential. Let us
stress that in the present case  the tadpole condition  defining the
bosonic open string    is not compulsory: the presence of the dilaton
tadpole does not  render the  theory  inconsistent if the vacuum is
shifted by the Fishler-Susskind mechanism \cite{fish}.

The tadpole condition for the 26-dimensional uncompactified bosonic
string determines the Chan-Paton group to be $SO(2^{13})$. Let us
discuss  the derivation of this well-known result. Introducing a
Chan-Paton multiplicity
$n$ at  both string ends, the four  different one-loop amplitudes of
the unoriented 26-dimensional bosonic string  are given  by (see for
example \cite{pob}):\begin{eqnarray}
\label{direct} {\cal T }=\int_{\cal F} {d^2\tau \over \tau_2^{14}}
{1\over
\eta^{24 }(\tau) {\bar \eta}^{24 }(\bar \tau)}\ , \\
\label{bottle}{\cal K } = {1\over2}\int_0^\infty {d\tau_2 \over
\tau_2^{14}} {1\over \eta^{24 }(2i\tau_2)}\ ,
\\ \label{annulus}{\cal A}={n^2\over2}\int_0^\infty {d\tau_2 \over
\tau_2^{14}}  {1\over
\eta^{24 }(i\tau_2/2)}\ , \\ \label{mob} {\cal M } =  {\epsilon~n
\over2}\int_0^\infty
 {d\tau_2 \over \tau_2^{14}}  {1 \over {\hat \eta}^{24 }(i\tau_2/ 2 +
1/2)}\ ,
\end{eqnarray} where $\cal F$ is a fundamental domain of the modular
group for the torus and $\eta(\tau)$ is the Dedekind function:
\begin{equation}
\eta(\tau) = q^{1\over 24} \prod_{m=1}^{\infty} (1-q^m),~~~~~~ 
q=e^{2\pi i
\tau},~~~~~\tau =\tau_1 +i \tau_2\ .
\end{equation} The `hatted' Dedekind function in ${\cal M}$ means that
the overall phase is dropped  in
$\eta(i\tau_2/ 2 + 1/2)$ ensuring that $\hat \eta(i\tau_2/ 2 + 1/2)$
is real. A similar notation will be used for a generic function $f$ 
admitting an expansion $f= q^a \sum_{i=0}^\infty a_i q^i$, namely
$\hat f(\tau + 1/2) = e^{-i\pi a} f(\tau + 1/2)$. The world-sheet
parity operator defining  $\cal M$ is
$\Omega = \epsilon\  (-1)^N $ where $N$ is the open string oscillator
number operator.   The plus or minus sign $\epsilon$   in
Eq.(\ref{mob}) encodes the action of
$\Omega$ on the vacuum and the one-half  shift in the argument  of
the Dedekind function in Eq.(\ref{mob}) encodes the action of the
twist operator $ (-1)^N $.

The amplitudes ${\cal T}/2 \ +\ {\cal K}$ and ${\cal A} \ +\ {\cal
M}$ are respectively  the partition function of the closed and open
unoriented string sectors.    The
$q$-independent term in the expansion of the integrand of ${\cal
A}+{\cal M}$    gives the number of massless vectors and determines
the nature of the Chan-Paton group. Using Eqs.(\ref{annulus}) and 
(\ref{mob}) one finds
$n( n -\epsilon  )/2$ massless vectors: if $\epsilon=+1$ (resp.
$\epsilon =-1$), the Chan-Paton group is
$SO(n)$ (resp.
$USp(n)$) .

To impose the tadpole condition we interpret  ${\cal K}$,  ${\cal A}$
and
${\cal M}$ as amplitudes  in the transverse  (tree) channel.  To this
effect we first define  $t=2 \tau_2$ (resp. $t=\tau_2 /2$) in
${\cal K}$ (resp. ${\cal A}$) and express the modular form
$\eta^{24}(it)$ in the integrand of Eqs. (\ref{bottle}) and
(\ref{annulus}) in terms of its
$S$-transform.  The change of variable $l=  1/t$ yields
\begin{eqnarray}
\label{trans1}
  {\cal K } _{tree} &=& {2^{13}\over2}\int_0^\infty dl {1\over
\eta^{24 }(il)}
 \ ,\\\label {trans1p}
\hbox{and}\qquad {\cal A } _{tree} &=&{n^2\
2^{-13}\over2}\int_0^\infty dl {1\over
\eta^{24 }(il)}\ .
\end{eqnarray} The subscript $tree$ emphasises that the expressions
Eqs.(\ref{trans1}) and  (\ref{trans1p}), although identical to the
integrals  Eqs.(\ref{bottle}) and (\ref{annulus}), are now rewritten
in terms of tree level intermediate states. Throughout the paper, any
amplitude
 formulated in the `direct' channel as a one loop amplitude $\cal I$,
will be relabeled ${\cal I}_{tree}$ when expressed in terms of
transverse channel tree level intermediate states.

It is a little bit more tricky to go from the direct to   the
transverse channel for the M{\"o}bius amplitude.  One   expresses
${\hat
\eta}^{24 }(i\tau_2/ 2 + 1/2)$ in terms of its $P$-transform
\cite{sagn1} which combines the modular transformations
$S$ (i.e.
$\tau
\rightarrow -1/ \tau$) and $T$ (i.e. $\tau
\rightarrow \tau +1$):
\begin{equation}
\label{ptra} P=T^{1/2}ST^2S T^{1/2}\ .
\end{equation} One then performs the change of variable $l=1/
(2\tau_2)$ to get
\begin{equation}
\label{trans2} {\cal M } _{tree} =  2{\epsilon~ n
\over2}\int_0^\infty dl {1\over
\hat\eta^{24 }(il + 1/2)}\ .
 \end{equation}

The tadpole condition can now be imposed by requiring the vanishing of
the  $e^{-2\pi l}$-independent term in the integrand of the  total
tree amplitude
$  {\cal K } _{tree}+  {\cal A } _{tree}+  {\cal M } _{tree}$.  One
gets the following condition
\begin{equation}
\label{not1} (2^{13}+2^{-13}n^2 - 2 \epsilon\  n)=2^{-13}(2^{13} -
\epsilon\ n)^2 =0\ ,
\end{equation} which singles out $\epsilon =+1$ and        the value
$n= 2^{13}$.

  Therefore,  one recovers that the uncompactified
 open bosonic string theory obeying the tadpole condition is
unoriented and has an $SO(2^{13})$ Chan-Paton group \cite{sob}.

\subsection{Compactification of the open string ancestor on Lie
algebra lattices and truncation}

We now explain how the open string theories in 10 dimensions  may be
obtained by truncation from the compactified 26-dimensional bosonic
string. We begin by  discussing the construction of open string
descendants from closed strings  compactified on the two EN lattices
of the rank sixteen groups ${\cal G} = E_8 \times E_8$ and 
${\cal G} = E_8 \times SO(16)$. We refer the reader 
to ref. \cite{eht}  for a more general discussion on the open descendants 
of strings  compactified on EN lattices of semi-simple 
Lie groups ${\cal G}$. Having the amplitudes, we then perform the truncation.

\subsubsection{The  $E_8 \times E_8$ compactification and Type I
superstring}

We consider here the  compactification  on the
$E_8 \times E_8$ lattice.  We write
\begin{equation}
\gamma_{{(o)}_{E_8}}
\gamma_{{(o)}_{E_8}}=(\gamma_{{(o)}_{16}} +
\gamma_{{(s)}_{16}}) (\gamma_{{(o)}_{16}} +
\gamma_{{(s)}_{16}})\ .
\end{equation}

In order  to get the  amplitudes in this case, 
one first replaces the contribution of sixteen non-compact dimensions 
 in the direct amplitudes eqs.(\ref{bottle}),
(\ref{annulus}) and (\ref{mob}) by the contribution of the 
lattice partition function and then using the modular transformations
one gets the tree amplitudes which  read \cite{eht}
\begin{eqnarray}
\label{ksup} {\cal K } _{tree}& =& {2^5 \over2}\int_0^\infty dl~~
{1\over
\eta^{8 }(il)}
\gamma_{{(o)}_{E_8}}(il) \gamma_{{(o)}_{E_8}}(il)\ ,
\\ \label{asup}{\cal A } _{tree}& =&{n^2\ 2^{-5}\over2}\int_0^\infty
dl~~ {1\over \eta^8 (il)}
\gamma_{{(o)}_{E_8}}(il) \gamma_{{(o)}_{E_8}}(il)\ ,\\ {\cal M }
_{tree}& =&
\epsilon n \int_0^\infty dl~~ {1 \over {\hat \eta}^8 (il+{1 \over
2})} {\hat
\gamma}_{{(o)}_{E_8}}(il+1/ 2)  {\hat \gamma}_{{(o)}_{E_8}}(il+ 1 /
2)\ .
\label{msup}
\end{eqnarray}

We       impose the vanishing  of the dilaton tadpole.  All the
transverse amplitudes are  proportional to the root lattice of 
$E_8\times E_8$. Therefore
{\em all} divergences  are
 eliminated by a
 single constraint
\begin{equation}
\label{not2}  n=2^5 \ ,
\end{equation} and $\epsilon = +1$. The Chan-Paton group  is then
$SO(2^5)=SO(32)$. This reduction of Chan-Paton group
from $SO(2^{13})$ by compactification is crucially related to the
structure of the Lie algebra lattice. Indeed the reduction $ 2^{13} 
\rightarrow 2^5$ in Eqs.(\ref{ksup}) and (\ref{asup})  is a result
of the reduction of the number of non-compact dimensions 
and of the modular properties of the lattice partition function 
of the $E_8 \times E_8$ lattice. The latter is  
 characterised by one conjugacy class and is thus invariant under the modular
group. If we would  compactify instead   on a 
cartesian torus corresponding to $[U(1)]^{16}$  no reduction would  occur.
Indeed, in  this case the S-transformation of 
the $[U(1)]^{16}$ lattice partition function would  compensate
the change due to the reduction of  non-compact dimensions.  

We are now in position to derive the truncated theory from the
tadpole-free
 bosonic open string theory  compactified on the EN lattice of $E_8
\times E_8$. Eq.(\ref{truncations}) gives
\begin{equation}
\gamma_{(o)_{E8}}\gamma_{(o)_{E8}} = 
\gamma_{(o)_{E8}}(\gamma_{(o)_{16}}+
\gamma_{(s)_{16}})\rightarrow \hbox{Truncation}\rightarrow
\gamma_{(v)_{8}}-\gamma_{(s)_{8}}\, 
\end{equation}

and the transverse amplitudes Eqs.(\ref{ksup}),
(\ref{asup}) and (\ref{msup}) become after truncation
\begin{eqnarray}
\label{Ktt} {\cal K }^{t} _{tree}& =& {\cal A}^{t}_{tree} =  {2^5
\over 2}\int_0^\infty dl~~ {1\over \eta^8(il) }
(\gamma_{(v)_{8}}-\gamma_{(s)_{8}})(il) \ ,
 \\ \label{Mtt}  {\cal M }^{t} _{tree}& =& - 2^5 \int_0^\infty dl~~ {1
\over {\hat
\eta}^8 (il +1 / 2) } (\hat \gamma_{(v)_{8}}  - \hat \gamma_{(s)_{8}})
 (il +1 / 2) \ .
\end{eqnarray} The flip in sign  in Eq.(\ref{Mtt}) as compared to
Eq.(\ref{msup})  with $\epsilon=+1$ is solely due to the definition
of hatted functions. Truncating the direct amplitudes ${\cal A}$ and
${\cal M}$ we get
\begin{eqnarray}
 \label{At} {\cal A}^{t}&=& {2^{10}\over2}\int_0^\infty {d\tau_2
\over \tau_2^6} {1\over \eta^8 (i\tau_2/ 2)} ( \gamma_{(v)_{8}} -
\gamma_{(s)_{8}})(i\tau_2/ 2)\ ,\\ \label{Mt} {\cal M}^t& =& - {2^5
\over 2} \int_0^\infty  {d\tau_2 \over \tau_2^6} {1\over {\hat
\eta}^8 (i\tau_2/ 2+1 /2)} (\hat \gamma_{(v)_{8}} -
\hat\gamma_{(s)_{8}})(i\tau_2/ 2+1 /2)\ .
 \end{eqnarray} The amplitudes in  Eqs.(\ref{At}) and (\ref{Mt}) are
equal to those in
 Eqs.(\ref{Ktt}) and (\ref{Mtt}) expressed in the transverse channel,
a
 consequence of the fact that the truncation commutes with $S$
and $T$
 transformations.

Massless modes in both open and closed string channels, created by
bosonic oscillators in non-compact dimensions, become massive after
truncation.  New massless modes arise from the compact dimensions.
 In the closed string  channel, these are  massless spinors, 
$NS$-$NS$ and
$R$-$R$ fields. The $NS$-$NS$ and $R$-$R$   tadpoles  are
 eliminated by the  condition Eq.(\ref{not2}) inherited from the
bosonic string. This is easily checked  in the truncated tree amplitude ${\cal K }^{t}
_{tree}+{\cal A }^{t} _{tree}+{\cal M }^{t} _{tree}$ given by   Eqs.(\ref{Ktt})
and (\ref{Mtt})     . 

The Chan-Paton group  $SO(32)$ is preserved under truncation. This
can be checked  by counting, in the open channel,  the number of
massless vectors
 arising from truncation. 

We see that the  truncation of the unoriented tadpole-free bosonic
open string theory, compactified on the
$E_8
\times E_8$  lattice,  results in  the $SO(32)$ anomaly free Type I
theory.

\subsubsection{The $E_8 \times SO(16)$ compactification and Type 0
strings}

We now consider  the 26-dimensional  unoriented bosonic open string
theory compactified on the $E_8 \times SO(16)$ lattice. In this case
there are {\it four} conjugacy classes ${\cal N}$ which give, compared to
Eq.(\ref{trans1}), a reduction $2^{13} \rightarrow 2^5 {\sqrt{\cal N}}= 2^6$. 
Taking into account the
existence of these four classes and the possibility of introducing
different Chan-Paton multiplicities, the tree amplitudes of this 
model are given by \cite{eht}

\begin{eqnarray}
\label{ESt} {\cal K } _{tree} &=& {2^6 \over2}\int_0^\infty dl {1\over
\eta^8 }
\gamma_{{(o)}_{16}} \gamma_{{(o)}_{E8}}\ ,
\\ \label{ASt}{\cal A } _{tree} &=&{2^{-6}\over2}\int_0^\infty dl
{1\over
\eta^8} (a^2_1 \gamma_{{(o)}_{16}} + a^2_2 \gamma_{{(v)}_{16}}  +
a^2_3
\gamma_{{(s)}_{16}}    + a^2_4 \gamma_{{(c)}_{16}})
\gamma_{{(o)}_{E_8}},\\
\label{MSt} {\cal M } _{tree} &=& \epsilon a_1
\int_0^\infty dl {1
\over {\hat \eta}^8}  {\hat \gamma}_{{(o)}_{16}}{\hat
\gamma}_{{(o)}_{E_8}}\ ,
\end{eqnarray} and
\begin{eqnarray}
\label{chanv} a_1&=&n_o+n_v+n_s+n_c\ , ~~~~~ a_2=n_o+n_v-n_s-n_c\ ,
\nonumber \\ a_3&=&n_o-n_v+n_s-n_c \ ,~~~~~ a_4=n_o-n_v-n_s+n_c\ .
\end{eqnarray}

We  enforce the tadpole conditions. In this case there are four
tadpole conditions because  there are four different types of
massless modes giving rise  to divergent contributions in the tree
amplitudes: in addition to the graviton and dilaton encoded in the
Dedekind function at level one, there are  three  types of massless modes
arising from
$\gamma_{(o)_{16}}$ at level 1 and from
$\gamma_{(s)_{16}}$,
$\gamma_{(c)_{16}}$ at level zero. The tadpole conditions which
eliminate the divergences at level one in the Dedekind function and in
$\gamma_{(o)_{16}}$ fixe $\epsilon =1$,  and the Chan-Paton group is
  $SO(n_o) \times SO(n_v) \times SO(n_s) \times SO(n_c)$, with
$n_o+n_v+n_s+n_c=2^6=64 $. The pattern of symmetry breaking of $SO(64)$
is further determined by imposing the two remaining tadpole
conditions corresponding to the
$\gamma_{(s)_{16}}$ and $\gamma_{(c)_{16}}$ divergences. One gets
\begin{equation}
\label{grob} SO(n) \times SO(n) \times SO(32-n) \times SO(32-n)\ .
\end{equation} 
We now derive the truncated theory.
Using Eq.(\ref{truncations}) we get
\begin{eqnarray}
\label{EStt} {\cal K }^t_{tree} &=& {2^6 \over2}\int_0^\infty dl
{1\over
\eta^8 }
\gamma_{{(v)}_8} \ ,
\\ \label{AStt}{\cal A }^t_{tree} &=&{2^{-6}\over2}\int_0^\infty dl
{1\over
\eta^8} [(64)^2 \gamma_{{(v)}_{8}} +
 16 (n-16)^2 \gamma_{{(o)}_{8}}] \ ,  \\ \label{KStt}{\cal M
}^t_{tree} &=& -64
\int_0^\infty dl {1 \over {\hat \eta}^8}  {\hat \gamma}_{{(v)}_{8}} \
,
\end{eqnarray} from which the direct channel amplitudes can be obtained. The
Chan-Paton group is transferred from the untruncated bosonic theory to the
truncated fermionic theory and  we obtain the spectrum of the tadpole-free
$[SO(32-n) \times SO(n)]^2$ Type O  theories discussed in
ref.\cite{sagn1,bega}.  Note that the Chan-Paton group has a higher rank  than
in Type I because $E_8 \times  SO(16)$ has more conjugacy classes than
$E_8\times E_8$. We emphasize that this Chan-Paton gauge structure
and the symmetry breaking pattern arise entirely from bosonic
considerations.

\subsection{Brane fusion}

We shall relate the Chan-Paton groups of the tadpole free bosonic string 
compactified on the enhanced symmetry points considered in the previous
section (and hence of the open fermionic strings) to the uncompactified
$SO(2^{13})$ unoriented  bosonic string.

The latter  has a clear
geometrical interpretation \cite{pol}.  The ends of the open strings
live on  D25-branes and the tension of a D25-brane can be derived from
${\cal A}_{tree}$  (see Eq.(\ref{trans1p}) with $n=1$) by comparing
with the field theory calculation \cite{pol}.

The tension $T^{bosonic}_{D25}$ is given by\footnote{We always give 
the Dp-brane tensions  computed in the oriented closed theory.}
\begin{equation}
\label{brane} T^{bosonic}_{D25} = {\sqrt\pi\over2^4
  \kappa_{26}}(2\pi\alpha^\prime{}^{1/2})^{-14}\ ,
\end{equation}
 where $\kappa_{26}^2= 8\pi G_{26}$ and $G_{26}$ is the Newtonian
constant in 26 dimensions.

The action of the world-sheet  parity operator on the 26-dimensional
closed bosonic string introduces  an orientifold 25-plane $O25$. The
tension of the $O25$ can be derived from ${\cal K}_{tree}$ (and ${\cal
M}_{tree}$ for the sign of this tension)  again by  comparing with
the field theory calculation. The result  is \cite{pol}:
\begin{equation}
\label{orientifold}  T_{O25}^{bosonic} = - 2^{12} T^{bosonic}_{D25}\ .
\end{equation} Therefore the tadpole condition fixing the
$SO(2^{13})$ gauge group  means in this context that one has to
introduce $n=2^{13}$ D25-branes (
$2^{12}$+ their images) to cancel the negative tension of  the $O25$
\cite{pol}.

We now derive the tension of the wrapped orientifold and D25-branes
for compactifications on the EN lattices with Lie group ${\cal G}=E_8 
\times E_8$ and
 ${\cal G}=E_8 \times SO(16)$.
To this effect we describe the lattice in terms of 
constant background metric $g_{ab}$ and Neveu-Schwarz antisymmetric
tensor $b_{ab}$ \cite{egrs}.

The  squared wrapped orientifold tension  is obtained from the
coupling to gravity in the transverse Klein bottle amplitude ${\cal
K}_{tree}$.  We recall that ${\cal
K}_{tree}$ is obtained from the direct  amplitude ${\cal K}$ by the S
modular transformation, and that
${\cal K}$ follows from the torus amplitude by inserting the
world-sheet parity operator $\Omega_c$ which interchanges left and
right sectors. This projects out the $b_{ab}$-field. For the
above groups, the tension of the wrapped orientifold $T_{O25~
wr}^{bosonic}$ after  compactification  is then given
by
\begin{equation}
\label{worien} T_{O25~ wr}^{bosonic}=T_{O25}^{bosonic} [(2 \pi)^{16}
\sqrt{g}]\ ,
\end{equation} where $g$ is the determinant of the metric. 
On the other hand the world-volume action of a  D25-brane is given 
by the Born-Infeld action \cite{tens1,tens2,cal}. Accordingly,  after
compactification,  the tension $T_{D25~ wr}^{bosonic}$ of a wrapped
D25-brane is given  by \cite{hkms}:
\begin{equation}
\label{wdbrane}  T_{D25~ wr}^{bosonic}=T_{D25}^{bosonic} [(2 \pi)^{16}
\sqrt{e}] = n_f T_{D25}^{bosonic} [(2 \pi)^{16}
\sqrt{g}]\ ,
\end{equation} where $e$ is the determinant of $e_{ab} \equiv
g_{ab}+b_{ab}$, and
\begin{equation}
\label{reduc} n_f = \sqrt{ e /g}\ .
\end{equation} The negative tension of the orientifold can  be
compensated   by introducing
$n_w$ wrapped D25 branes (including images) with:
\begin{equation}
\label{crit} n_w =  2^{13} \sqrt{g \over e} = {2^{13}\over n_f} \ .
\end{equation} The reduction  factor $n_f$  defined in
Eq.(\ref{reduc}) can be  computed \cite{eht} to give
\begin{equation}
\label{fuse}  n_f ={2^{8} \over \sqrt {\cal N}}\,
\end{equation} 
where $ {\cal N}$ is the number of conjugacy classes\footnote{One can use
Eqs.(\ref{wdbrane}), (\ref{fuse}) and (\ref{brane})  to compute the tensions
of   branes in the {\em fermionic} string theories by comparing the zero mode
contributions in 
${\cal A }_{tree}$  and ${\cal A }^t_{tree}$ both in the  $E_8 \times E_8$ and 
$E_8
\times SO(16)$ compactifications \cite{eht}. The results obtained in this way
agree with the known results.}

Inserting this value in Eq.(\ref{crit}), we get the number of  wrapped D25
branes required to ensure the  dilaton tadpole condition in the
compactified bosonic string. Thus the  reduction factor
$n_f$ coincides with the ratio of Chan-Paton multiplicities before
and after compactification.

For $E_8 \times E_8$ we have ${\cal N}=1$ and  Eq.(\ref{fuse}) gives
$n_f = 2^8$. Consequently we are left with
$n_w=2^5$
 wrapped D25-branes.  This is consistent with  the fact that the
Chan-Paton group of the tadpole-free
$E_8 \times E_8$ compactification is  $SO(32)$.

For $E_8 \times SO(16)$ we have  ${\cal N}=4$. Thus Eq.(\ref{fuse})
gives
$n_f = 2^7$. Consequently we are left with
$n_w=2^6$ wrapped  D25-branes. This is consistent with  the fact that the
Chan-Paton group of the tadpole-free
$E_8 \times SO(16)$ compactification is  $[SO(32-n) \times SO(n)]^2$.

Note that, for the two Lie groups of rank $d=16$ considered here one has 
\begin{equation}
\label{rank} {2^{d/2}\over\sqrt{\cal N}}= 2^{r/2} \ ,
\end{equation} 
where $r$ is the rank of the $b_{ab}$ matrix.
This can be verified by direct inspection from the
explicit form of $b_{ab}$.
Such reduction, and its expression in terms of the rank of the
$b_{ab}$ matrix,   is in agreement with reference
\cite{sagn2}.

One may interpret the reduction of the total Chan-Paton multiplicity
as a consequence of  `brane fusion'.  Eq.(\ref{wdbrane}) shows that
{\em one} D25-brane of the 26-dimensional bosonic string theory,
compactified on the EN lattice,  has the same tension as the integer number
$n_f=\sqrt{e/g}$ of D25-branes when the  theory is compactified on a
generic Cartesian torus (for which
$b_{ab}=0$) with  equal volume. This fact does not depend on the
orientability of the string nor on the tadpole condition. In presence
of the enhanced symmetry, the  r\^ole of the quantised $b_{ab}$-field
is  to  fuse
$n_f$ component branes into {\em one} .

\section{Discussions}

One  may  always bosonise the world-sheet fermions to
obtain, in the light-cone gauge,  theories formally identical to
those resulting from truncation\footnote{See
\cite{revlsw} and references therein.}.  The added information
provided by truncation  for closed strings is the encoding of  all these
theories   in the 26-dimensional bosonic string theory compactified,  in the
left and/or right sector, on the Lie algebra lattice of $E_8\times
SO(16)$.  

The extension of the truncation to open string sectors reveals that this
encoding explains, in terms only of properties of the bosonic strings, crucial
properties of the fermionic strings, such as  the cancellation of anomalies in
Type I theory by the Chan-Paton group $SO(32)$, and  the Chan-Paton groups
which eliminate all tadpoles in Type O theories.  Perhaps more
surprising even is the fact that these results follow from specific
properties of the $E_8\times SO(16)$ lattice, singling out this lattice out of 
all possible toroidal compactifications.

At this stage, the  link between all M-theory strings and the bosonic string
results mainly from properties of  conformal symmetry and of group
theory.  Namely, the preservation of modular invariance by truncation and
the fusion mechanism on enhanced symmetry points play a key role in
deriving our results.  One is tempted to inquire further into the nature of
this link and to search for  its possible non perturbative dynamical origin. The
emergence of the space-time fermions from diagonal subgroups of direct
products of space-time groups and  internal groups is,  as mentioned in the
introduction, suggestive of the generation of fermions through excitations
from topological boson backgrounds \cite{thooft, min}.  In this context, the
group
$E_8$ (and/or $SO(16)$) might play a key r\^ole in attempting to
construct some ``bosonic M-theory''~\cite{suss}.

\end{document}